\documentclass[prl,onecolumn]{revtex4}
\usepackage{amsmath}
\usepackage{amssymb}
\usepackage{mathptmx}
\usepackage{ifpdf}
\usepackage[pdftex]{graphicx}
\DeclareGraphicsExtensions{.eps, .jpg, .pdf, .tif}
\usepackage[pdftex]{hyperref}
\usepackage[percent]{overpic}
\newcommand{\be}{\begin{equation}}
\newcommand{\ee}{\end{equation}}
\newcommand{\ber}{\begin{eqnarray}}
\newcommand{\eer}{\end{eqnarray}}
\def\vx{{\bf x}}
\def\vy{{\bf y}}
\def\vz{{\bf z}}
\def\va{{\bf a}}
\def\hx{\hat{x}}
\def\hy{\hat{y}}
\def\hz{\hat{z}}
\newcommand{\hvx}{\hat\vx}
\newcommand{\hvy}{\hat\vy}

\newcommand{\He}{$^3$He}
\newcommand{\Hea}{$^3$He-A}
\newcommand{\upt}{UPt$_{3}$}
\newcommand{\sro}{Sr$_{2}$RuO$_{4}$}
\newcommand{\hvp}{\hat\vp}
\newcommand{\eps}{\varepsilon}
\newcommand{\sgn}{\mbox{sgn}}
\renewcommand{\Im}{\mbox{Im\,}}
\renewcommand{\Re}{\mbox{Re\,}}
\def\ZtwoOrbit{{\tt Z_{\text{2}}^{\text{orbit}}}}
\def\TwoDparity{{\tt P}_{2}}

\def\parity{{\tt P}}
\def\time{{\tt T}}
\def\point#1#2{{\tt #1}_{\mbox{\footnotesize #2}}}
\def\vr{{\bf r}}
\def\vv{{\bf v}}
\def\vp{{\bf p}}
\def\vq{{\bf q}}
\def\root{\sqrt{2}}
\def\ns{\negthickspace}

\def\abs#1{\vert #1\vert}
\def\cA{{\mathcal A}}
\def\cL{{\cal L}}
\def\cU{{\cal U}}
\def\cK{{\cal K}}
\def\cF{{\cal F}}
\def\vA{{\bf A}}
\def\Aem{{\bf {\mathbb A}}}
\def\J{{\bf {\mathbb J}}}
\def\K{{\bf {\mathbb K}}}
\def\E{{\bf {\mathbb E}}}

\def\nicefrac#1#2{\frac{#1}{#2}}
\begin{document}
\title{
\vspace*{-20mm}
\hspace*{-10cm}{\small Published in Frontiers in Physics 3:36 (2015)
[\href{http://journal.frontiersin.org/article/10.3389/fphy.2015.00036/abstract}{doi: 10.3389/fphy.2015.00036}]
}\\
\vspace*{5mm}
Anisotropy and Strong-Coupling Effects on the Collective Mode Spectrum of Chiral 
Superconductors: Application to \sro\footnote{Invited Talk at SCES14, July 7-11, 2014, 
Grenoble, France}
}

\author{J. A. Sauls$^{\dag}$, Hao Wu$^{\dag}$, Suk Bum Chung$^{\flat}$} 
\address{$^{\dag}$Department of Physics and Astronomy, 
             Northwestern University, Evanston, IL 60208 USA}
\email{sauls@northwestern.edu}
\address{$^{\flat}$Center for Correlated Electron Systems, 
				   Institute for Basic Science}
\address{$^{\flat}$Department of Physics and Astronomy, 
				   Seoul National University, Seoul 151-747, Korea}
\date{February 27, 2015; revised May 23, 2015}

\begin{abstract}
Recent theories of \sro\ based on the interplay of strong interactions, spin-orbit coupling and
multi-band anisotropy predict chiral or helical ground states with strong anisotropy of the
pairing states, with deep minima in the excitation gap, as well as strong phase
anisotropy for the chiral ground state.
We develop time-dependent mean field theory to calculate the Bosonic spectrum for the class of
2D chiral superconductors spanning \Hea\ to chiral superconductors with strong anisotropy.
Chiral superconductors support a pair of massive Bosonic excitations of the time-reversed pairs 
labeled by their parity under charge conjugation. These modes are degenerate for 2D \Hea.
Crystal field anisotropy lifts the degeneracy. 
Strong anisotropy also leads to low-lying Fermions, and thus to channels for the decay of the
Bosonic modes.
Selection rules and phase space considerations lead to large asymmetries in the lifetimes and
hybridization of the Bosonic modes with the continuum of un-bound Fermion pairs. 
We also highlight results for the excitation of the Bosonic modes by microwave radiation 
that provide clear signatures of the Bosonic modes of an anisotropic chiral ground state. 
\end{abstract}
\maketitle

\vspace*{-3mm}
\section*{Introduction}
\vspace*{-3mm}

Superfluid \He\ and unconventional superconductors share a common and fundamental property that the
ground state breaks one or more symmetries of the normal Fermionic vacuum in conjunction with the
usual $\point{U(1)}{\ns}$ gauge symmetry associated with BCS condensation.
In the case of \sro\ the connection with superfluid \He\ may be stronger. 
The theoretical proposal for a spin-triplet, p-wave ground state in \sro\ was motivated 
by similarities between \sro\ and liquid \He, particularly the existence of exchange enhanced
paramagnetism in a strongly correlated Fermi liquid \cite{ric95}.
In liquid \He\ long-lived ferromagnetic spin fluctuations (``paramagnons'') are believed to
be the mechanism responsible for the BCS pairing instability to a spin-triplet, p-wave ground state
\cite{lay71,leg75}.
The Balian-Werthamer (BW) state \cite{bal63}, identified as the B-phase, with total angular momentum
$J=0$ fully gaps the 3D Fermi surface, and as a result minimizes the free energy in the
weak-coupling limit. The BW state is time-reversal invariant, but spontaneously 
breaks relative spin-orbit rotation symmetry.
However, spin-triplet correlations, which differentiate between the Anderson-Morel (AM)
state and BW states, feedback to modify the spin-fluctuation exchange interaction,
leading to stabilization of the AM state at high
pressures \cite{bri73,bri74}. This is the chiral A-phase which breaks time-reversal ($\time$)
symmetry and reflection symmetry in any plane containing the chiral axis ($\TwoDparity$), but
preserves $\time\times\TwoDparity$ (chiral symmetry).

Similarly, Rice and Sigrist argued that for \sro, which is a layered perovskite belonging to the
$\point{D}{4h}$ point group, the likely pairing state, in analogy with \He, would be the planar
state, the 2D analog of the BW state.
In weak-coupling theory, and neglecting spin-orbit coupling, the planar state, which
belongs to the one-dimensional $\point{A}{1u}$ representation of $\point{D}{4h}$, is degenerate with
the 2D chiral AM state, which belongs to the 2D $\point{E}{1u}$ representation. Thus, if
spin-fluctuation exchange is also the mechanism for pairing in \sro, we expect the strong-coupling
feedback effect will stabilize the chiral AM state.
However, in addition to the near two-dimensionality of the Fermi surface of \sro, incommensurate
spin-density-wave (SDW) fluctuations connected with the quasi-one-dimensional $\alpha$ and $\beta$
bands co-exist with ferromagnetic fluctuations at lower temperatures \cite{bra02}. Spin-orbit
coupling, and the possibility of pairing on multiple Fermi surface sheets likely play important
roles in determining the pairing symmetry class, ground state order parameter
\cite{hav08,rag10,wan13,sca14,vee14} as well as the Bosonic excitation spectrum in \sro.

In what follows we develop a field theory for the Bosonic modes based on the Ginzburg-Landau theory
(TDGL) for spin-triplet, odd-parity pairing that allows one to compare predictions for \He\ and
\sro, and to examine the roles of crystalline anisotropy and strong correlation effects on the
Bosonic modes.
The Bosonic modes are excitations of a condensate of Cooper pairs for which the parent state is the
Fermionic vacuum. Thus, although the TDGL field theory provides insight into the
Bosonic excitation spectrum, it misses key features of a more complete theory of the low-lying
excitations of the BCS pair condensate. Notably, 
(i) polarization effects of the underlying Fermionic vacuum on the excitation energies (masses and
dispersion) of the Bosonic modes,
(ii) finite lifetimes of the Bosonic excitations due to coupling to the Fermionic continuum and
(iii) selection rules and matrix elements for the coupling of the Bosonic modes to the
electromagnetic (EM) field.
Key predictions of a microscopic theory of interacting Fermionic and Bosonics modes are summarized,
including microwave spectroscopic signatures of the Bosonic excitation spectrum.

\vspace*{-3mm}
\subsection*{Order Parameter}
\vspace*{-3mm}

The mean-field order parameter for spin-triplet Cooper pairs,
$\Delta_{\alpha\beta}(\vp)=d_{\mu}(\vp)(i\sigma_{\mu}\sigma_y)_{\alpha\beta}$, where $\alpha,\beta$
label the projections of fermion spins of the Cooper pair,
$\{i\sigma_{\mu}\sigma_y\,|\,\mu=x',y',z'\}$ is the spin-triplet basis of $2\times 2$ matrices, is
expressed as
\be
d_{\mu}(\vp) = \sum_{i=x,y,z}\,A_{\mu i}\,(\hvp_i)\,,\quad \mu\in\{x',y',z'\}
\,,
\ee
where $d_{\mu}(\vp)$ is a vector under rotations in spin space, and is a function of the vector
basis of p-wave orbital basis functions, $(\hvp_x,\hvp_y,\hvp_z)$, for bulk \He.
Thus, the amplitudes $A_{\mu i}$ provide a bi-vector representation of
$\point{SO(3)}{S}\times\point{SO(3)}{L}$.
For quasi-two-dimensional \sro, the orbital basis, $(Y_x(\vp),Y_y(\vp))$, provides a 2D vector
representation in which $Y_{x,y}(\vp)$ transforms as $\vp_{x,y}$ under the point group
$\point{D}{4h}$. These basis functions reflect anisotropic pairing on the Fermi surface.

The chiral AM state is the stable equilibrium phase in a narrow temperature window 
$T_{\text{AB}}\le T < T_c$ of bulk \He\ at pressures $P\ge P_c = 21\,\mbox{bar}$. 
However, if we confine \He\ as a thin film or within a thin cavity of thickness $D \le D_c\approx
1\,\mu\mbox{m}$ the quasi-2D A-phase with
$\vec{d}(\vp) = \nicefrac{\Delta}{\sqrt{2}}\,\,\hz\,\left(\hvp_x \pm i \hvp_y\right)$
is the ground state for all pressures and temperatures \cite{vor07}. We refer to this order
parameter as the `Anderson-Morel (AM) state', the `chiral state' or the `A-phase order parameter'.

However, in the weak-coupling limit the chiral AM phase is degenerate with the 2D planar phase,
$\vec{d}(\vp) = \nicefrac{\Delta}{\sqrt{2}}\,\,\left(\hx\,\hvp_x + \hy\,\hvp_y\right)$. 
In the context of \sro\ the 2D planar state is referred to as the ``helical state'' \cite{sca14}.
The degeneracy between the chiral and helical ground states is lifted by strong-coupling
corrections to weak-coupling theory, or spin-orbit coupling.
For \He\ strong-coupling effects dominate the nuclear spin-orbit coupling, with the
spin-triplet feedback effect on the ferromagnetic spin-fluctuation exchange interaction favoring the
chiral AM state over the helical state. 
Recent NMR experiments on \He\ confined in thin slabs imply that the chiral AM state is stable
relative to the helical state down to $P\approx 0\,\mbox{bar}$ \cite{lev13}.
If superconductivity in \sro\ is also driven by
ferromagnetic spin-fluctuations then we expect the chiral phase to be favored.
However, recent theoretical calculations based on RG analysis that includes spin-orbit coupling
within a multi-band pairing model also find both helical and chiral phases, depending upon the
interaction parameters defining the lattice pairing Hamiltonian \cite{sca14}.
For the chiral state the pairing gap, $|\Delta(\vp)|\equiv|\vec{d}(\vp)|$, is 
found to be strongly anisotropic on all three bands with deep gap minima
over a wide range of material parameters \cite{sca14}.
Thus, the main observations are: 
(i) there are competing low energy scales that determine the ground state for \sro, 
(ii) it is not currently settled whether or not \sro\ is a chiral superconductor \cite{kal09}, or
even if the order parameter for \sro\ belongs to a two-dimensional representation \cite{hic14}, and
(iii) whatever the ground state - e.g. chiral or helical - low-energy Bosonic excitations may
provide unique signatures of the ground state based on their symmetry and selection rules.

\vspace*{-3mm}
\section*{Ginzburg-Landau Theory}
\vspace*{-3mm}

Consider the Ginzburg-Landau theory for the class of equal-spin pairing (ESP) states of confined
superfluid \He\ in 2D. A simple generalization also describes spin-triplet superconductivity on a 2D
cylindrical Fermi surface, i.e. pairing on the $\gamma$ band in \sro\ within the $\point{E}{1u}$
representation of $\point{D}{4h}$.
The general form of the order parameter is then given by
\be
\vec{d}(\vp) = \hat{d}\,\left(A_x\,Y_x(\vp) + A_y\,Y_y(\vp) \right)
\,,
\ee 
where $\{Y_i(\vp) | i = x,y\}$ are basis functions for the 2D irreducible representation,
$\point{E}{1u}$, of $\point{D}{4h}$, and $\vA=A_{x}\hvx + A_{y}\hvy$ is a complex vector describing
pairing in this generalized ``p-wave'' orbital basis.
In what follows we consider the class of ESP states in which the direction $\hat{d}$ along which the
Cooper pairs have zero spin projection is fixed, either as a spontaneously broken symmetry direction,
or by spin-orbit coupling.
The general form of the GL free energy functional is constructed from invariants of the maximal
symmetry group from products of $A_i$ and $A_i^*$ through 4th order \cite{hes90},
\be\label{eq-GL_functional_E1u}
\cF[\vA] 
	= 
\int_{\text{V}}\,dV\,\Big\{\alpha(T)\,\vert\vA\vert^2
						 + \beta_1\,\vert\vA\vert^4 + \beta_2\,\vert\vA\cdot\vA\vert^2
					     + \beta_3\,\big[\vert A_x\vert^4 + \vert A_y\vert^4 \big]
					  \Big\}
\,,
\ee
where $\alpha(T)$ determines the onset of pairing in the $\point{E}{1u}$ representation, i.e.
$\alpha(T_c)\equiv 0$, and thus, $\alpha(T)\approx\alpha'(T-T_c)$. The fourth order coefficients,
$\beta_{1,2,3}$, determine the magnitude of the condensation energy density and the relative
stability of phases belonging to the $\point{E}{1u}$ symmetry class.
The first three terms in Eq. \ref{eq-GL_functional_E1u} are invariant under the larger group, i.e.
$\point{U(1)}{\negthickspace}\times\point{SO(2)}{$L_z$}\times\ZtwoOrbit\times\parity\times\time$,
while the last term is an additional invariant that is allowed in the case of \sro\ by the lower
symmetry, $\point{U(1)}{\negthickspace}\times\point{D}{4h}$.

\begin{figure}[t]
\hspace*{-8.5cm}\begin{overpic}[scale=0.60]{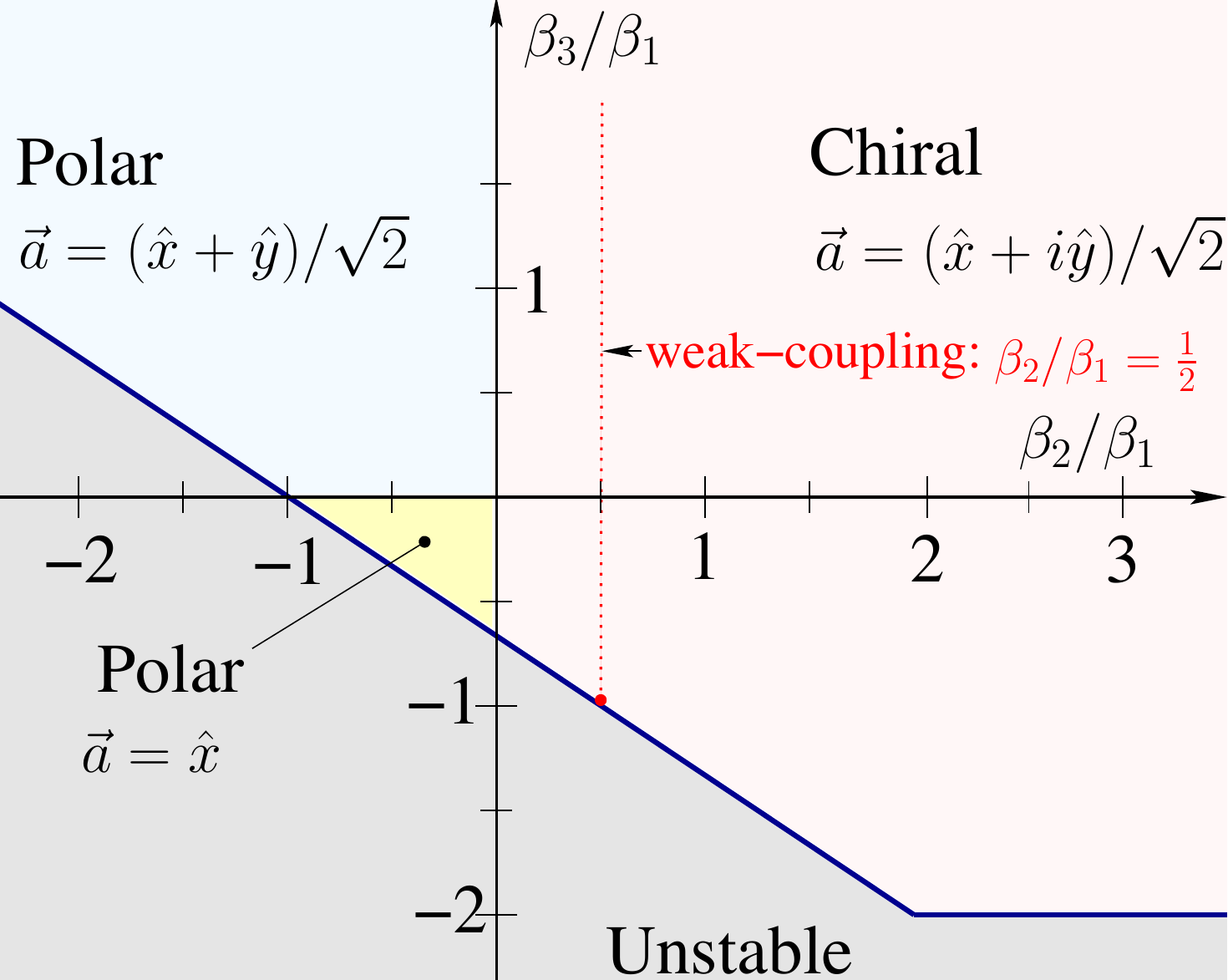}
\put(105,-5){\includegraphics[scale=0.55]{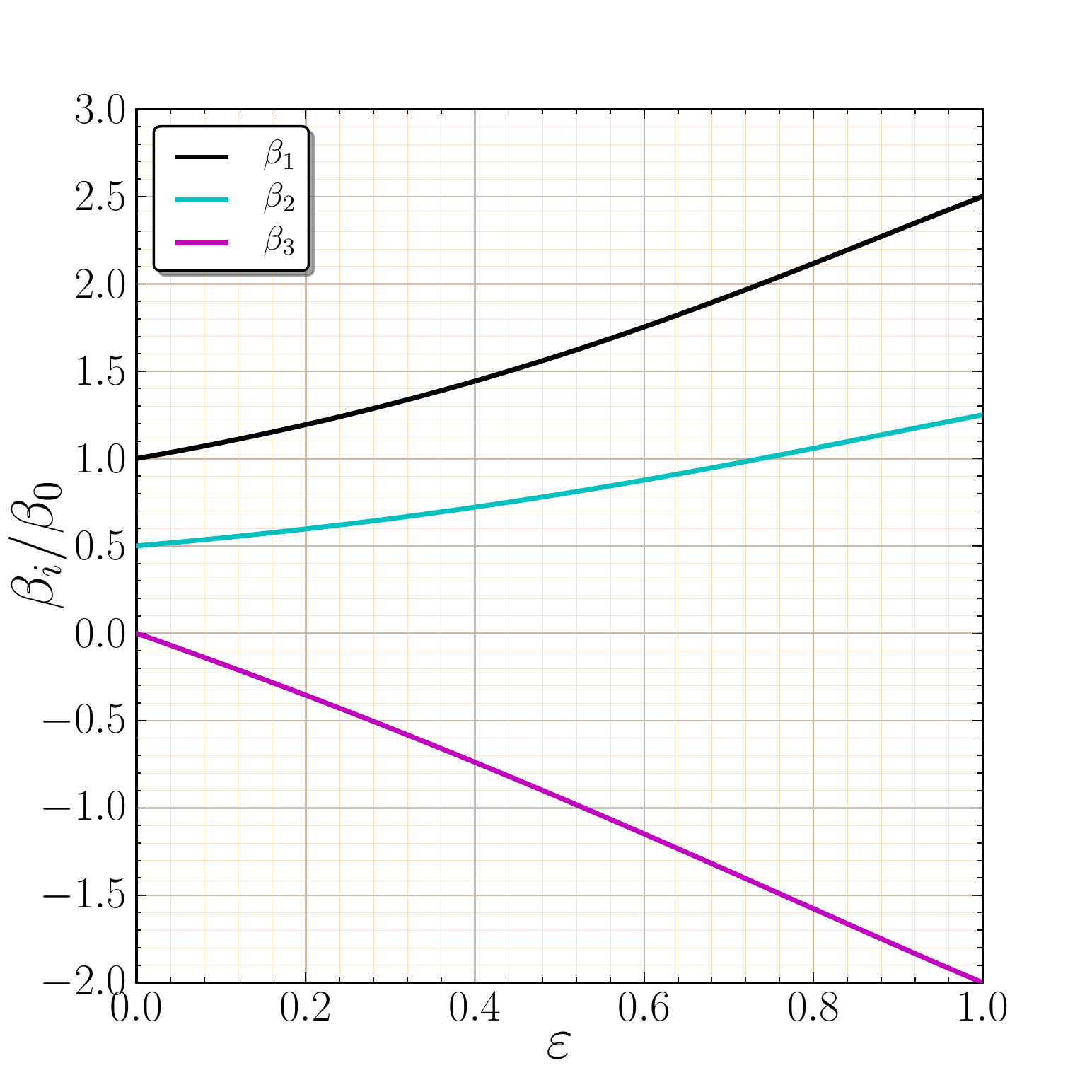}}
\end{overpic}
\caption{
	Left: Ginzburg-Landau phase diagram for $\point{E}{1u}$ pairing. 
	The weak-coupling BCS theory predicts the
	$\beta$ parameters lie on the dotted red line depending on the degree of anisotropy of the
	$\point{E}{1u}$ basis functions, implying the chiral AM state is stable.
 	Right: Anisotropy of the GL $\beta$ parameters in the weak-coupling limit
    based on the anisotropic pairing model defined in Eq. \ref{eq-E1u_basis} as a function 
	of the anisotropy parameter $\epsilon$ calculated from Eqs. 
	\ref{eq-beta1_weak-coupling} - \ref{eq-beta3_weak-coupling}.
} 
\label{fig-phase_diagram_E1u}
\end{figure}

It is convenient to express the bulk order parameter in terms of an amplitude and a normalized
complex vector, $\vA = \Delta\,\va$, with $\vert\va\vert^2\equiv\va\cdot\va^*=1$. The normalized
order parameter is then parametrized by two angles,
$\va=\left(\cos\vartheta\hvx+e^{i\varphi}\sin\vartheta\hvy\right)/\sqrt{2}$
The resulting GL functional is then
\be\label{eq-GL_functional_E1u-2}
\cF[\Delta,\va]
	=\int_{\text{V}}dV\,\big\{\alpha(T)\Delta^2 + \tilde\beta[\va]\,\Delta^4\big\}
\,,\,\mbox{with}\quad
	\tilde\beta[\va]\equiv\beta_1+\beta_2\vert\va\cdot\va\vert^2
	+\beta_3[\vert a_x\vert^4+\vert a_y\vert^4]
\,.
\ee
For $T<T_c$, minimization with respect to $\Delta$ leads to $\Delta_{\text{min}}^2
=-\alpha(T)/2\tilde\beta[\va]$, and a condensation energy given by
$\cF[\va]=-\alpha(T)^2/4\tilde\beta[\va]$, with $\tilde\beta[\va] > 0$ for global stability.
The ground state is then determined by the normalized order parameter that minimizes
$\tilde\beta[\va]=\beta_1+\beta_2(1-\sin^2\varphi\sin^2(2\vartheta))+\beta_3(1-\nicefrac{1}{2}\sin^2
(2\vartheta))$.
In the case of 2D \He\ we have $\beta_3 = 0$, in which case there are two possible ESP ground
states; for $\beta_1 > 0$ and $\beta_2 > 0$ the chiral AM state which breaks time-reversal and 2D
parity is preferred. There are two degenerate chiral ground states which are time-reversed partners
of one another, $\va_{\pm}=(\hvx\pm i\hvy)/\sqrt{2}$.
However, for $-\beta_1 < \beta_2 < 0$ the in-plane polar state with
$\va=\cos\vartheta\hvx+\sin\vartheta\hvy\,,\mbox{for}\,,\vartheta\in [0,2\pi]$ minimizes the GL
functional. This phase has a continuous degeneracy with respect to orientation of polar
axis in the $x-y$ plane.
Tetragonal anisotropy lifts the continuous degeneracy of the in-plane polar state. 
For $\beta_2 < 0$ and $\beta_3>0$ the polar state aligns along a [110] direction, e.g.
$\va=(\hvx+\hvy)/\sqrt{2}$. However, for $-\nicefrac{2}{3}(1+\beta_2/\beta_1) < \beta_3/\beta_1 < 0$
the polar state aligns along a [100] direction, e.g. $\va=\hvx$.
The phase diagram is shown in Fig. \ref{fig-phase_diagram_E1u}. Note that the weak-coupling
prediction for the $\beta$ parameters favor a chiral ground state independent of the measure of
anisotropy, i.e. $\beta_3$. Furthermore, substantial corrections to weak-coupling theory are
required to stabilize an in-plane polar state. In fact the helical state, which belongs to a
different irreducible representation ($\point{A}{1u}$), is a more likely competitor to the chiral
state since these two states are degenerate in the weak-coupling limit without spin-orbit coupling.
In addition, AFM spin-fluctuations can lead to attractive, sub-dominant, pairing
interactions in even parity, ``d-wave'', channels.
Here we consider fluctuations within the $\point{E}{1u}$ representation, and neglect possible 
low-lying fluctuations of the ``helical'' phase ($\point{A}{1u}$) or even-parity,
$\point{B}{1g}$ or $\point{B}{2g}$, ``d-wave'' states.

\vspace*{-3mm}
\section*{Time Dependent GL Theory - Fluctuations}
\vspace*{-3mm}

Consider the Bosonic excitations of the chiral AM ground state with $\va_{+} =
(\hvx+i\hvy)/\sqrt{2}$.
Space-time fluctuations of the $\point{E}{1u}$ order parameter, $\cA(\vr,t)=\vA - \Delta\va_{+}$,
are represented by two complex amplitudes,
\be\label{eq-OP_fluctuations}
\cA(\vr,t) = D(\vr,t)\,\va_{+} + E(\vr,t)\,\va_{-}
\,.
\ee
There are two classes of excitations - modes with chirality $L_z = +1$ represented by the
field $D(\vr,t)$ and modes with the time-reversed chirality, $L_z=-1$, represented by the field
$E(\vr,t)$ - and altogether four orbital collective modes within the $\point{E}{1u}$ representation.
We construct an effective Lagrangian by expanding the GL free energy functional
about the ground state. Time-dependent fluctuations introduce an additional invariant,
$\cK = \int_{\text{V}}dV\,\nicefrac{1}{2}\mu\,\dot\cA_{i}\dot\cA_{i}^{*}$, where $\mu$ is
the effective inertia of the Cooper pair fluctuations and $\dot{\cA}=\partial_t\cA$.
The effective potential is obtained by expanding the GL functional to 2$^{nd}$ order in the
fluctuations $\cA(\vr,t)$: $\cU[\cA] = \cF[\vA]-\cF[\Delta\va_{+}]$.
The Lagrangian, $\cL=\cK-\cU$, takes a simplified form when expressed in amplitudes for the
normal modes: $D^{\pm}=D\pm D^{*}$ and $E^{\pm} = E\pm E^{*}$,
\be
\cL
= \int_{\text{V}}dV\,
\Bigg\{
	\nicefrac{1}{2}\mu\,\Big[(\dot{D}^{+})^2 + (\dot{D}^{-})^2 
	                        + (\dot{E}^{+})^2 + (\dot{E}^{-})^2\Big]
	-4\Delta^2\Big[\beta_1 (D^{+})^2 + \beta_2((E^{+})^2 + (E^{-})^2) 
					+ \nicefrac{1}{2}\beta_3((D^{+})^2 + (E^{+})^2)\Big]
\Bigg\} 
\,.
\ee
The Euler-Lagrange equations reduce to four uncoupled mode equations,
\be
\ddot{D}^{\text{C}} + M_{\text{+ C}}^2\,D^{\text{C}} = 0
\quad\mbox{and}\quad
\ddot{E}^{\text{C}} + M_{\text{- C}}^2\,E^{\text{C}} = 0
\,,
\ee
where $M_{\text{L$_z$ C}}$ is the excitation gap (``mass'') of the Bosonic mode with quantum numbers
$L_z,C$, and $C$ is the parity under charge conjugation (particle-hole) symmetry
\cite{ser83a,fis85}.
The amplitude $D^{-}$, which is a cyclic coordinate in the Lagrangian, is a Goldstone mode
associated with the broken $\point{U(1)}{\negthickspace}$ gauge symmetry. Indeed for
small amplitude and phase fluctuations of the chiral ground state,
$\vA(\vr,t)=\Delta(\vr,t)\,e^{i\theta(\vr,t)}\,\va_{+}$ 
$\approx \Delta\,\va_{+} (1 + \delta(\vr,t) + i\theta(\vr,t))$, we identify 
$D^{-}=i\Delta\,\theta(\vr,t)$ with the phase fluctuation and
$D^{+}=\Delta\,\delta(\vr,t)$ with the amplitude fluctuation. The amplitude mode of the ground-state
order parameter is the Higgs mode for the chiral ground state. In particular, $D^{+}$
corresponds to an excitation with the same quantum numbers ($L_z=+1$ and $C=+1$) as those of 
the ground state.
In the weak-coupling limit the mass of the amplitude mode is determined the pair-breaking energy for
dissociation into two Fermions, i.e. $M_{\text{++}}=2 m_{\text{F}}$ where $m_{\text{F}}=\Delta$ is
the gap (mass) in the Fermionic spectrum of the broken symmetry phase \cite{hig64,lit81}.
Furthermore, since the amplitude mode has the same quantum numbers as the condensate, the mass of
the amplitude mode is unshifted, relative to that of two dissociated Fermions, by polarization
effects of the underlying Fermionic vacuum. 
We use this to fix the inertia term in the effective Lagrangian as $\mu=(2\beta_1+\beta_3)$, and
thus the mass scale for all the Bosonic modes of the effective Lagrangian.

\begin{table}[t]
\begin{center}
\begin{tabular}{c|c|c|c|c}
		Mode	&	Symmetry		&	Mass		&	Name		&	EM active
\\
\hline
	$D^{+}$		&	$L_z=+1$ $C=+1$	&	$2\Delta$	&	Amplitude 	&	no	
\\
	$D^{-}$		&	$L_z=+1$ $C=-1$	&	$0$			& 	Phase Mode	&	no	
\\
	$E^{+}$		&	$L_z=-1$ $C=+1$	&	$\root\Delta$	
													&	E$^{+}$ Mode &	yes	
\\
	$E^{-}$		&	$L_z=-1$ $C=-1$	&	$\root\Delta$			
													& 	E$^{-}$ Mode&	yes	
\\
\hline
\end{tabular}
\end{center}
\caption{Bosonic Mode Spectrum for the 2D Chiral Ground State $\va_{+}$. 
The masses of the $E^{\pm}$ modes are those for an isotropic 2D chiral condensate 
in the weak-coupling limit. Also indicated is whether or not the mode 
can be excited by absorption of microwave photons.}
\label{table-Mode_Spectrum}
\end{table}

\begin{figure}[t]
\begin{center}
\hspace*{-7mm}
\includegraphics[width=0.50\hsize]{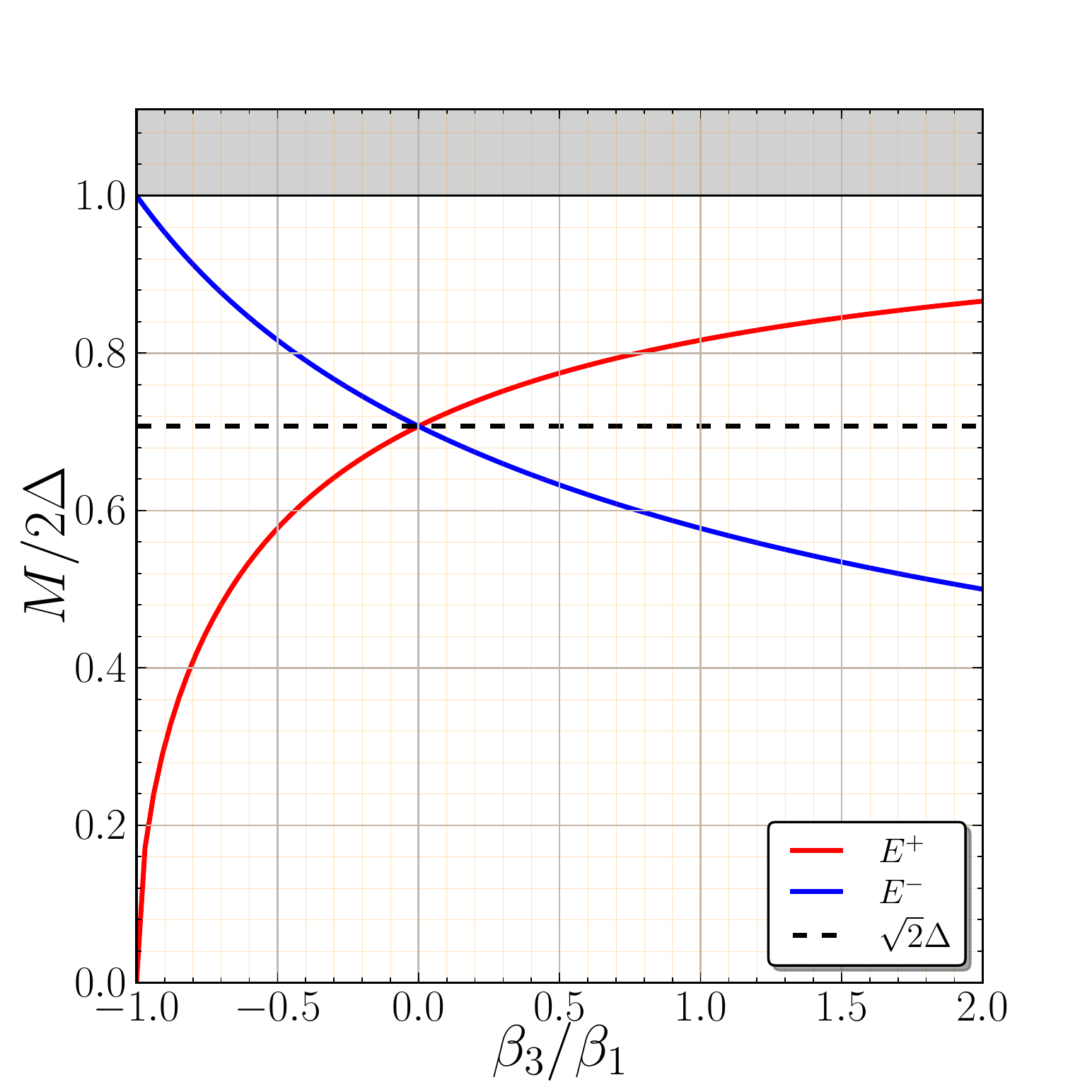}
\includegraphics[width=0.50\hsize]{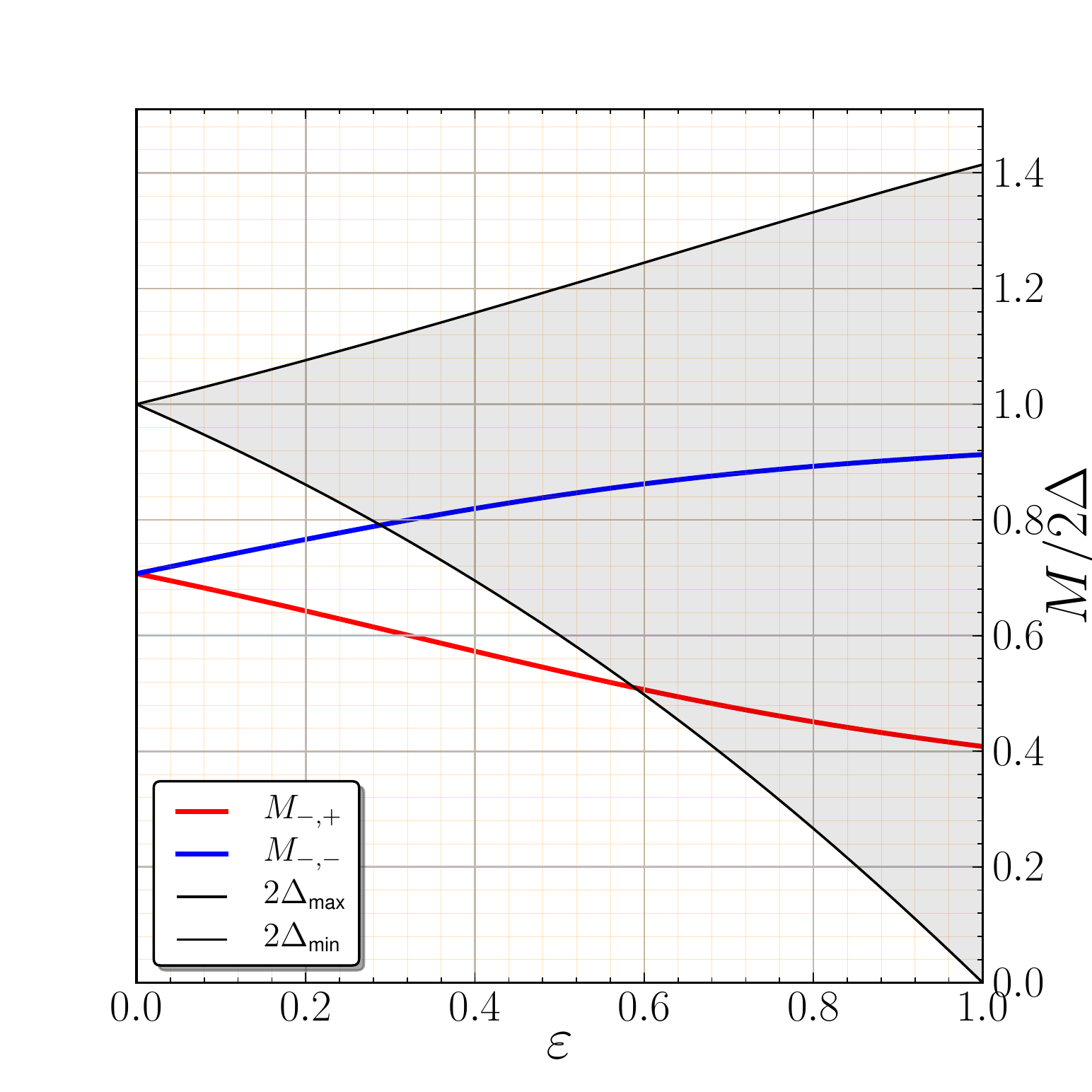}
\caption{Left: Masses of the $E_{\pm}$ modes versus the GL anisotropy parameter, $\beta_3/\beta_1$, 
		in the weak-coupling limit, $\beta_2/\beta_1 = \nicefrac{1}{2}$. 
		Right: $E_{\pm}$ masses based on the weak-coupling $\beta$ parameters 
		defined in Eqs. \ref{eq-beta1_weak-coupling}-\ref{eq-beta3_weak-coupling}
		for the anisotropic $\point{E}{1u}$ basis functions defined in Eq. \ref{eq-E1u_basis}.
		The maximum and minimum of the pair-breaking edge, $2|\Delta(\vp)|$, 
		bound the gray shaded region.
}
\label{fig-AH_Mode_Splitting}
\end{center}
\end{figure}

Collective modes of the Cooper pair condensate with quantum numbers distinct from those of the
ground state were discussed soon after the formulation of BCS theory by Anderson \cite{and58},
Bogoliubov, Shirkov and Tolmachev \cite{bogoliubov58}, Tsuneto \cite{tsu60}, Vaks, Galitskii and
Larkin \cite{vak61}, Bardasis and Schrieffer \cite{bar61}, and Vdovin \cite{vdo63}.
Generalizations of the amplitude mode were discovered theoretically in the context of superfluid
\He\ by Maki \cite{mak74}, W\"olfle \cite{wol74}, Sauls and Serene \cite{sau81}.
The observation of these Bosonic modes using acoustic spectroscopy played an important role in
identifying the symmetries of the superfluid phases of \He\ \cite{hal90,sau00a}.
In this context the collective modes for $\point{E}{1u}$ pairing symmetry with the time-reversed
chirality ($L_z=-1$), both of which are massive, correspond to the ``clapping modes'' of superfluid
\Hea\ in the weak-coupling limit of an isotropic 2D chiral AM state (See Table
III in Ref. \cite{wol74}). In particular, the masses of the $E^{\pm}$ modes in the effective
Lagrangian are given by
\ber
M_{\text{- +}} = 2\Delta\,\left(\frac{2\beta_2+\beta_3}{2\beta_1+\beta_3}\right)^{\nicefrac{1}{2}}
		\xrightarrow[\beta_2/\beta_1=\nicefrac{1}{2}]{\beta_3=0}\sqrt{2}\Delta
		\,,\quad (E^{+}\,\mbox{mode})
		\,,
\\
M_{\text{- -}} = 2\Delta\,\left(\frac{2\beta_2}{2\beta_1+\beta_3}\right)^{\nicefrac{1}{2}}
		\xrightarrow[\beta_2/\beta_1=\nicefrac{1}{2}]{\beta_3=0}\sqrt{2}\Delta
		\,,\quad (E^{-}\,\mbox{mode})
		\,.
\eer
Note that in the weak-coupling limit ($\beta_2/\beta_1=\nicefrac{1}{2}$) for an isotropic
($\beta_3=0$) 2D chiral ground state the time-reversed modes are \emph{degenerate}, and lie well
below the Fermionic continuum edge at $2\Delta$.
However, the degeneracy of the $E^{\pm}$ modes is lifted by tetragonal anisotropy of the Fermi
surface and pairing basis functions, which leads to $\beta_3\ne 0$. The crystal field splitting of
the $E^{\pm}$ modes is shown in Fig. \ref{fig-AH_Mode_Splitting}.
In the left panel the splitting of the modes in the weak-coupling limit
($\beta_2/\beta_1=\nicefrac{1}{2}$) is plotted as a function of $\beta_3/\beta_1$.
The soft $E^{+}$ mode, i.e. $M_{\text{- +}}\rightarrow 0$ for $\beta_3\rightarrow -2\beta_2$, is the
dynamical signature of the boundary of the unstable region of the GL phase diagram shown in Fig.
\ref{fig-phase_diagram_E1u}.

\vspace*{-3mm}
\section*{Weak-coupling GL Theory for Anisotropic $\point{E}{1u}$ pairing}
\vspace*{-3mm}

For quantitative predictions of the effects of anisotropy of the Fermi surface and
pairing interaction on the collective mode spectrum we require the angle-resolved density of states
on the Fermi surface, $n(\vp)$, and the anisotropy of the pairing basis functions,
$\{Y_x(\vp),Y_y(\vp)\}$, that define the $\point{E}{1u}$ orbital order parameter in momentum
space, $\Delta(\vp)=A_x\,Y_x(\vp) + A_y\,Y_y(\vp)$.
For a single-band Fermi surface the GL functional for ESP states in the weak-coupling limit can be
expressed in terms of Fermi-surface averages of the mean-field order parameter defined on the Fermi
surface \cite{sau94a,ali11},
\be\label{eq-GL_Functional_wc}
\cF_{\text{wc}} 
= 
\int_{\text{V}}dV\,\Big\{
\alpha(T)\,
\Big\langle\vert\Delta(\vp)\vert^2\Big\rangle
+
\beta_{0}\,
\Big\langle\vert\Delta(\vp)\vert^4\Big\rangle
\Big\}
\,,
\ee
where $\alpha(T)=\ln(T/T_c)$, $\beta_{0}=7\zeta(3)N_f/(4\pi T_c)^2$, $N_f$ is the single-spin
density of states at the Fermi energy, and
$\langle\ldots\rangle\equiv\int\,d^2p\,n(\vp)(\ldots)$ is the angle-average over the Fermi
surface.
The pairing basis functions are normalized with respect to the anisotropic Fermi surface,
$\langle Y^{*}_i(\vp)Y_j(\vp)\rangle = \delta_{ij}$ for
$i,j\in\{x,y\}$.
Evaluating the angular averages in Eq. \ref{eq-GL_Functional_wc} gives the following results 
for the GL material parameters in the weak-coupling limit,
\ber
\beta_{1}^{\text{wc}} 
&=& 
2\beta_{0}\,\Big\langle \vert Y_x(\vp)\vert^2 \vert Y_y(\vp)\vert^2 \Big\rangle
\,,
\label{eq-beta1_weak-coupling}
\\
\beta_{2}^{\text{wc}} 
&=& 
\beta_{0}\,\Big\langle \vert Y_x(\vp)\vert^2 \vert Y_y(\vp)\vert^2 \Big\rangle
\,,
\label{eq-beta2_weak-coupling}
\\
\beta_{3}^{\text{wc}} 
&=& 
\beta_{0}\,
\Big\langle \vert Y_x(\vp)\vert^4 - 3 
            \vert Y_x(\vp)\vert^2 \vert Y_y(\vp)\vert^2 \Big\rangle
\label{eq-beta3_weak-coupling}
\,.
\eer
These results are obtained by using the transformation properties of the basis functions under 
the group operations of $\point{D}{4h}$: specifically,
$Y_x\xrightarrow[]{C_4}Y_y$, $Y_y\xrightarrow[]{C_4}-Y_x$, 
$Y_x\xrightarrow[]{\Pi_{yz}}-Y_x$, $Y_y\xrightarrow[]{\Pi_{yz}}Y_y$.
A key result is that the ratio of GL parameters,
$\beta_{2}^{\text{wc}}/\beta_{1}^{\text{wc}}=\nicefrac{1}{2}$, is \emph{independent of anisotropy}
\cite{hes90}. Thus, the chiral AM ground state with broken time-reversal symmetry is favored
independent of the anisotropy of the $\point{E}{1u}$ basis functions and Fermi surface anisotropy in
the weak-coupling limit.

\vspace*{-3mm}
\subsection*{Cylindrical Symmetry}
\vspace*{-3mm}

For cylindrical symmetry, e.g. a Fermi disk for 2D \Hea, we have $n(\vp)=\frac{1}{2\pi p_f}$ and
$d^2p = p_f\,d\phi$. The normalized p-wave basis functions are then
$Y_x=\sqrt{2}\hvp_x=\sqrt{2}\cos\phi$ and $Y_y=\sqrt{2}\hvp_y=\sqrt{2}\sin\phi$, where $\phi$ is the
azimuthal angle defining the unit vector $\hvp$ normal to the edge of the Fermi disk.
Thus, for the chiral ground state the excitation gap (Fermion mass), 
$|\Delta(\vp)|\equiv\nicefrac{\Delta}{2}\vert Y_x(\vp)+iY_y(\vp)\vert\equiv\Delta$, is also 
isotropic. 
These basis functions also lead to the following results for the $\point{E}{1u}$ GL $\beta$
parameters, $\beta_1=\beta_{0}$, $\beta_2=\nicefrac{1}{2}\beta_{0}$, and $\beta_3=0$, the latter as
expected for an isotropic Fermi surface with pure p-wave basis functions. These values give the
weak-coupling, isotropic results for the $E^{\pm}$ modes reported in Table
\ref{table-Mode_Spectrum}, i.e. the degenerate ``clapping'' modes of 2D \Hea\ with
$M_{\text{-,+}}=M_{\text{-,-}}=\sqrt{2}\Delta$. Note that the Nambu-Sum rule,
$\sum_{C}M_{L_z,C}^2=(2\Delta)^2$, is obeyed for both classes ($L_z=\pm 1$) of Bosonic modes for 2D
\Hea\ in the weak-coupling limit \cite{vol14}.

\vspace*{-3mm}
\subsection*{Phase Anisotropy}
\vspace*{-3mm}

Recent calculations of pairing instabilities for odd-parity pairing in \sro\ starting from
lattice models based d-band electrons and holes with on-site and near neighbor Hubbard interactions
and spin-orbit coupling lead to a helical or chiral ground state, but with \emph{strong
anisotropy}.
For the chiral state obtained from pairing on hybridized, quasi-1D $\alpha$ and $\beta$ bands
\cite{rag10}, the phase of $\Delta(\vp)$ changes rapidly upon crossing the $[110]$ planes,
leading to a low-energy collective mode \cite{chu12}.
To illustrate the effects of strong phase anisotropy on the collective modes consider an extreme
limit in which the $\point{E}{1u}$ basis functions are constant within each quadrant of the Fermi
surface, but change sign abruptly upon crossing any of the [110] planes, i.e. $Y_x=\sgn(\hvp_x -
\hvp_y)$ and $Y_y=\sgn(\hvp_y + \hvp_x)$. These basis functions have constant amplitude, but are
discontinuous in phase ($\pm\pi$) across the [110] planes.
They give the following results for the $\point{E}{1u}$ GL $\beta$ parameters, $\beta_1=2\beta_{0}$,
$\beta_2=\beta_{0}$, and $\beta_3=-2\beta_{0}$. For the $E^{\pm}$ modes this leads $M_{\text{-
-}}\rightarrow 2\Delta$ and $M_{\text{- +}}\rightarrow 0$. The soft $E^{+}$ mode is a dynamical
signature of the approach to the unstable region of the GL phase diagram for $\beta_3\rightarrow
-2\beta_2$ as shown in Fig. \ref{fig-phase_diagram_E1u}.

\vspace*{-3mm}
\subsection*{Amplitude Anisotropy}
\vspace*{-3mm}

\begin{figure}[t]
\includegraphics[width=0.49\hsize]{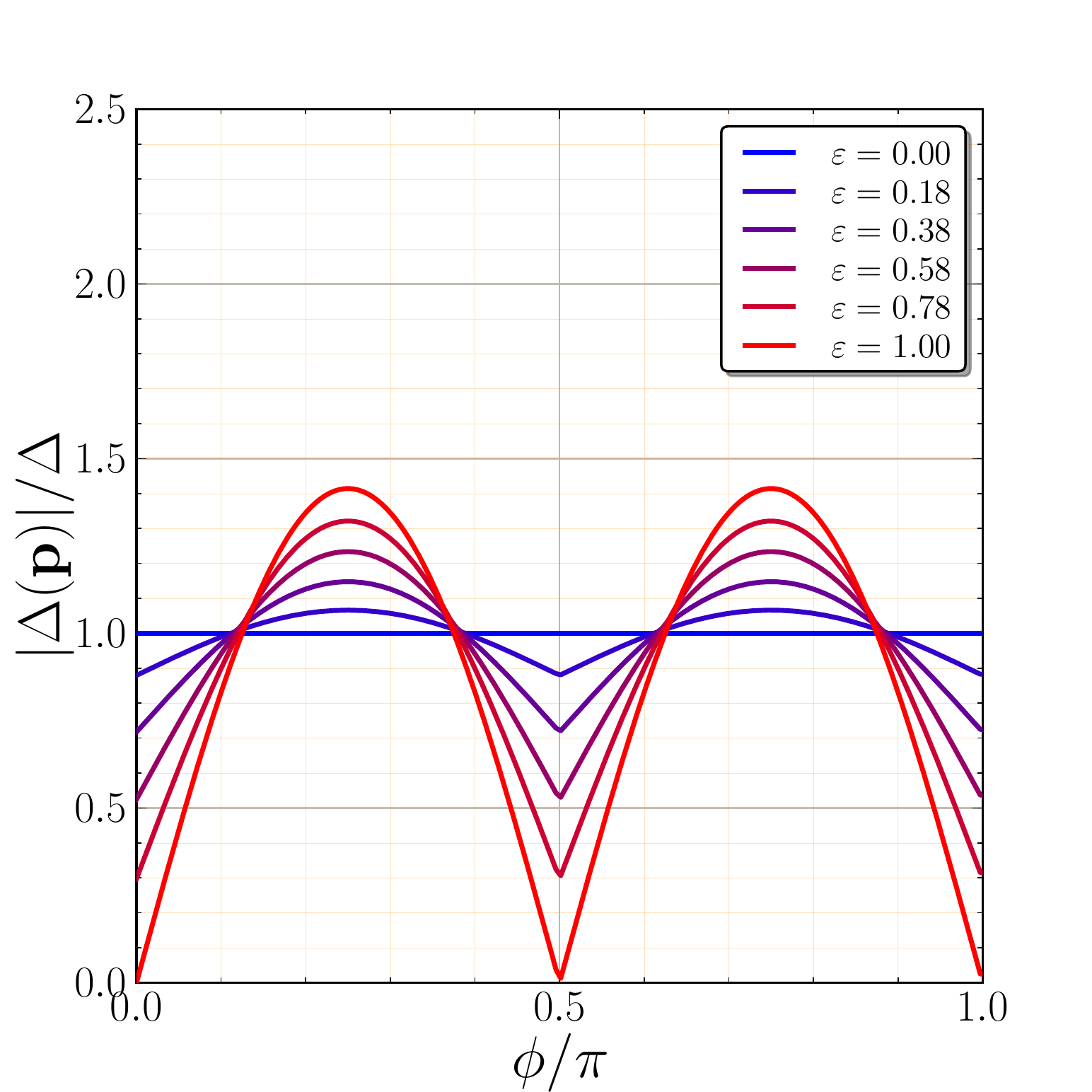}
\includegraphics[width=0.49\hsize]{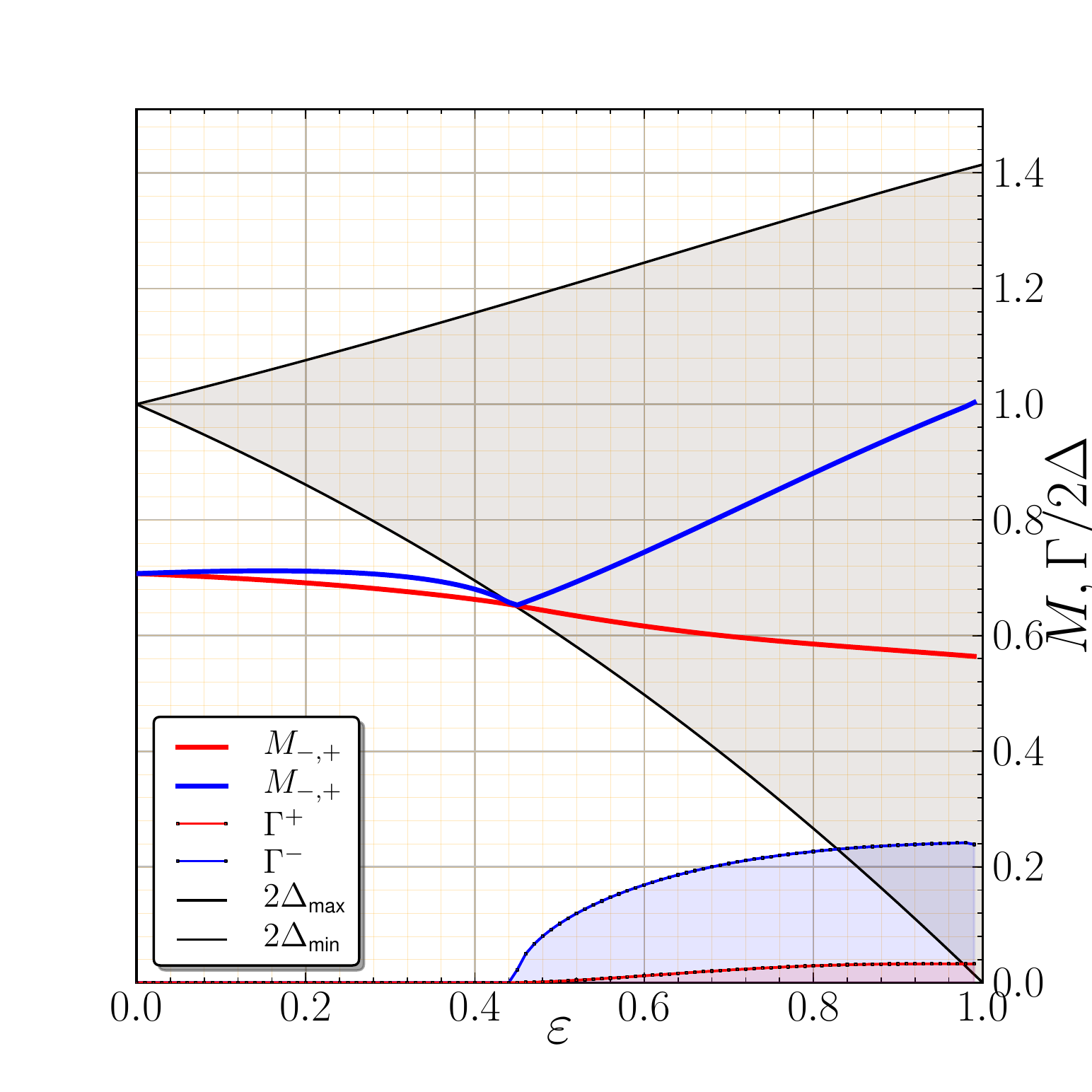}
\caption{
Left: Anisotropy of the excitation gap, $|\Delta(\vp)|$, based on 
			   Eqs. \ref{eq-E1u_basis}, 
			   exhibiting deep gap minima along [110] directions
			   in the limit $\epsilon\rightarrow 1$. 
Right: Masses and Linewidths of the $E^{\pm}$ modes resulting from gap anisotropy calculated from 
the Eqs. \ref{eq-Eplus} and \ref{eq-Eminus} at $T=0$.
} 
\label{fig-Gap-EMode_Anisotropy}
\end{figure}

Multi-band models leading to a chiral superconducting state also exhibit strong
anisotropy of the amplitude of the order parameter, in particular, an excitation gap $|\Delta(\vp)|$
with nodal, or near nodal, directions on the Fermi surface \cite{wan13,sca14}.
Strong anisotropy of the Cooper pair amplitudes in momentum space leads to physical properties that
are quite distinct from the predictions based on cylindrical symmetry of the $\gamma$ band of \sro,
including the splitting of the $E^{\pm}$ modes, and in some cases a low-energy collective mode.
To illustrate the effects of anisotropy of the pairing interaction on the collective mode spectrum,
as well as signatures of a chiral ground state with strong anisotropy, we consider the following
model for the anisotropic $\point{E}{1u}$ basis functions defined on the $\gamma$ band, or
hybridized $\alpha$ and $\beta$ bands,
\ber
Y_{x,y}(\vp) =  \sqrt{2}\,\hvp_{x,y}\,I(\vp)\,,\,\mbox{with}\,\,
I(\vp)       = \left(1+\epsilon(\vert 2\hvp_x\hvp_y\vert-1)\right)/
				\left(1 + 4\epsilon(1-\epsilon)/\pi
                    - 2\epsilon(1-3\epsilon/4)\right)^{\nicefrac{1}{2}}
\,,
\quad
0 \le \epsilon\le 1
\,.
\label{eq-E1u_basis}
\eer
Tetragonal anisotropy is parametrized by the function $I(\vp)$, which is invariant under
$\point{D}{4h}$. The $\point{E}{1u}$ basis functions are normalized,
$\int\frac{d\varphi}{2\pi}\,Y_i^*(\vp)Y_j(\vp) = \delta_{ij}$, and in the limit 
$\epsilon\rightarrow 1$ exhibit strong anisotropy with a minimum excitation gap,
$\Delta_{\text{min}}\propto(1-\epsilon)\rightarrow 0$ as $\epsilon\rightarrow 1$ in the $[100]$
directions.
The anisotropy of the excitation gap for the chiral ground state is shown in the left panel of 
Fig. \ref{fig-Gap-EMode_Anisotropy}, while the corresponding $\beta$ parameters calculated from 
Eqs. \ref{eq-beta1_weak-coupling} - \ref{eq-beta3_weak-coupling} are shown in the right panel of Fig. \ref{fig-phase_diagram_E1u}. 
The amplitude anisotropy has no effect on the ratio, $\beta_{2}/\beta_{1}$, that determines the
stability of the chiral state, but has a strong effect on the ratio, $\beta_3/\beta_1$, that
determines the effective potential for the $E^{\pm}$ modes, and thus the splitting of the these
modes by anisotropy, as shown in the right panel of Fig. \ref{fig-AH_Mode_Splitting}.
Note that in addition to the splitting of the $E^{\pm}$ modes the masses of the $E^{\pm}$ modes
cross the continuum edge ($2\Delta_{\text{min}}$) of unbound Fermion pairs. Thus, we expect the
$E^{\pm}$ modes to become resonances with finite lifetimes for sufficiently strong gap anisotropy.
However, the theory for the lifetimes of the $E^{\pm}$ modes is outside the TDGL Lagrangian
for the Bosonic spectrum, and requires a microscopic theory of the correlated Fermionic vacuum,
including the mechanism and effects of spontaneous symmetry breaking, and most importantly the
interaction and coupling between the Fermionic and Bosonic excitations of the chiral superconducting
phase.

\vspace*{-3mm}
\section*{Beyond TDGL}
\vspace*{-3mm}

A microscopic formulation of the electrodynamics of the excitations of p-wave superconductors,
including the coupling of Bosonic modes to a transverse (EM) electromagnetic field, is developed in
Refs. \cite{hir89,yip92} for 3D isotropic Fermi systems with p-wave, spin-triplet
pairing.
For a 3D chiral p-wave superconductor in the clean limit the coupled set of linearized dynamical
equations for the Bosonic mode spectra, including the reaction of the Fermionic vacuum to the
excitation of Bosonic modes, as well as the coupling of Bosonic and Fermionic excitations to the EM
field are formulated in Ref. \cite{yip92}.
We have extended this theory to 2D chiral superconductors with anisotropic, quasi-2D Fermi surfaces,
multi-band pairing and weak disorder, to make predictions for signatures of anisotropic chiral and
helical superconductivity based on the Bosonic mode spectrum and the microwave response for recent
theoretical models for the superconducting state of \sro\ \cite{wu15}.
Here we summarize some of the results from the microscopic theory that reflect the coupling between 
Bosonic and Fermionic degrees of freedom that are beyond the TDGL Lagrangian dynamics for chiral ground states.

The dynamics of the order parameter is formulated in terms of the space- and time-dependent mean-field pairing self-energy, which for spin-triplet fluctuations is given by
\be
\vec{d}(\vp;\vr,t) = 
\int d^2\vp' V(\vp,\vp')
\int\frac{d\varepsilon}{4\pi i}\vec{f}^{K}(\vp',\varepsilon;\vr,t)
\,,
\ee
where $V(\vp,\vp')=V_1\left(Y_{+}^*(\vp)Y_{+}(\vp')+Y_{-}^*(\vp)Y_{-}(\vp')\right)$ is the pairing 
interaction in the spin-triplet $\point{E}{1u}$ channel,
$\vec{f}^{\text{K}}(\vp,\varepsilon;\vr,t)$ is the anomalous Keldysh pair-propagator, the energy
integration is over the bandwidth of attraction near the Fermi level, $-\Omega_c\le\varepsilon\le
+\Omega_c$ with $\Omega_c\ll E_f$, and the momentum integration is an average over the Fermi surface
defined by the Fermi momentum $\vp$.
Below we discuss the orbital dynamics for an ESP chiral ground state of the form
$\vec{d}(\vp)=\hat{d}\,\Delta(\vp)$, with $\Delta(\vp)\equiv\Delta\,Y_{+}(\vp)$, with
$Y_{\pm}(\vp)\equiv\left(Y_x(\vp) \pm i Y_y(\vp\right)/\sqrt{2}$, and the spin component of the
order parameter, $\hat{d}$, is fixed along the direction $\hat\vz$.
The orbital fluctuations of the Cooper pairs are represented by two complex fields, 
\be
d(\vp;\vr,t) = D(\vr,t)\,Y_{+}(\vp) + E(\vr,t)\,Y_{-}(\vp)
\,,
\ee
where the notation is equivalent to that in Eq. \ref{eq-OP_fluctuations} of the TDGL theory. 
The solution to the linearized Eilenberger equations for the non-equilibrium pair-propagator,
$\vec{f}^{\text{K}}(\vp,\varepsilon;\vr,t)$, in terms of the time-dependent order parameter,
$\vec{d}(\vp;\vr,t)=\hat{d}\,d(\vp;\vr,t)$, and the coupling of charge currents to the EM field,
$\frac{e}{c}\vv_{\vp}\cdot\Aem(\vr,t)$ leads to coupled ``gap equations'' for the
orbital fluctuations of the Cooper pairs \cite{wu15}, 
\ber
\label{eq-gap_equation_d}
d(\vp;\vq,\omega)
\ns
&\ns\ns=\ns\ns&
\frac{1}{2}\int d^2\vp'\, V_t(\vp,\vp') 
	\left\{-\frac{1}{2}\bar{\lambda}(\vp')\,\eta'\,\Delta(\vp') 
	\left[\frac{2e}{c}\vv_{\vp}\cdot\Aem(\vq,\omega) \right]
	\right. 
\\
&\ns\ns+\ns\ns&\ns
	\left.
	\left[\gamma(\vp')
	+\frac{1}{2}\bar{\lambda}(\vp')(\omega^2-2|\Delta(\vp')|^2-\eta'^2)\right]\,
	 d(\vp';\vq,\omega)
	-\bar{\lambda}(\vp')
	\Delta(\vp')^2\,d'(\vp';\vq,\omega) 
	\right\}
	\,,
\nonumber
\\
\label{eq-gap_equation_dp}
d'(\vp;\vq,\omega)
\ns
&\ns\ns=\ns\ns&
\frac{1}{2}\int d^2\vp'\, V_t(\vp,\vp') 
	\left\{+\frac{1}{2}\bar{\lambda}(\vp')\,\eta'\,\Delta^*(\vp') 
	\left[\frac{2e}{c}\vv_{\vp}\cdot\Aem(\vq,\omega) \right]
	\right. 
\\
&\ns\ns+\ns\ns&\ns
	\left.
	\left[\gamma(\vp')
	+\frac{1}{2}\bar{\lambda}(\vp')(\omega^2-2|\Delta(\vp')|^2-\eta'^2)\right]\,
	 d'(\vp';\vq,\omega)
	-\bar{\lambda}(\vp')
	\Delta^*(\vp')^2\,d(\vp';\vq,\omega) 
	\right\}
	\,,
\nonumber
\eer
where $d'(\vp;\vr,t)\equiv D^*(\vr,t)\,Y_{+}(\vp) + E^*(\vr,t)\,Y_{-}(\vp)$,
$\eta'\equiv\vv_{\vp'}\cdot\vq$ generates the dispersion of both Fermionic and Bosonic excitations,
and we have expressed the gap equations in terms of Fourier modes.
Note in particular that the cross-coupling terms between $d(\vp;\vq,\omega)$ and
$d'(\vp;\vq,\omega)$ are proportional to the complex amplitudes, $\Delta(\vp)^2$ and
$\Delta^*(\vp)^2$.
The Tsuneto function \cite{mck90a},
\ber
\hspace*{-10mm}\bar\lambda(\vp;\omega,\vq)
&\equiv&
\int_{-\infty}^\infty\frac{d\varepsilon}{2\pi i} 
  \frac{\tanh\left(\displaystyle{\beta\abs{\varepsilon}/2}\right)}
       {\sqrt{\varepsilon^2 - \abs{\Delta(\vp)}^2}}\,
  \Theta (\varepsilon^2 - \abs{\Delta(\vp)}^2)
\nonumber
\\
&\times&
\left\{
\frac{\eta^2 + 2\omega(\varepsilon-\omega/2)}
     {\left(4(\varepsilon-\omega/2)^2-\eta^2\right)
      \left(\omega^2 - \eta^2 \right) + 4 \eta^2 \abs{\Delta(\vp)}^2} 
+
\frac{\eta^2 - 2\omega(\varepsilon+\omega/2)}
     {\left(4(\varepsilon+\omega/2)^2-\eta^2\right)
      \left(\omega^2 - \eta^2 \right) + 4 \eta^2 \abs{\Delta(\vp)}^2} 
\right\}
\,,
\hspace*{7mm}
\eer
determines 
(i) the coupling of the EM field to the Bosonic modes, 
(ii) the mass shifts for the Bosonic modes $D_{\pm}$ and $E^{\pm}$, 
(iii) finite lifetimes of Bosonic modes arising from coupling to the un-bound continuum, 
      i.e. when $M\ge 2\,\mbox{min}[|\Delta(\vp)|]$, and 
(iv) coupling of the EM field to the Fermionic spectrum, including absorption of 
	 EM radiation by creation of unbound Fermionic quasiparticles for
	 $\hbar\omega\ge 2|\Delta(\vp)|$.
Finally, the term $\frac{1}{2}\gamma(\vp)$ is the log-divergent integral that determines the BCS
instability and equilbrium gap function, $\Delta(\vp)$. Thus, we can regulate the divergence and
eliminate the pairing interaction, $V_1$, in the dynamical equations for $d(\vp)$ and $d'(\vp)$ in
favor of the self-consistently determined equilibrium gap function, $\Delta(\vp)$, using the
identity, $\frac{V_1}{2}\int d^2\vp\,Y_{\mu}^{*}(\vp)\,\gamma(\vp)\,Y_{\nu}(\vp) =
\delta_{\mu,\nu}$ for $\mu,\nu = \pm$.

\vspace*{-3mm}
\subsection*{Energies and Lifetimes of the $E^{\pm}$ modes}
\vspace*{-3mm}

The Bosonic modes are obtained from the eigenvalue spectrum of the homogeneous equations, i.e. for
$\Aem=0$. In the $\vq=0$ limit the eigen-modes are the linear combinations $D^{\pm}=D(\omega)\pm
D^*(-\omega)$ and $E^{\pm}=E(\omega)\pm E^*(-\omega)$ as in the TDGL theory, with $D^{-}$
representing the phase mode and $D^{+}$ the corresponding amplitude mode. For the modes with
time-reversed chirality we obtain
\ber
\label{eq-Eplus}
\left(\lambda_{00}(\omega)\,\omega^2 - 4 \Delta^2\,\lambda_{11}(\omega)\right) E^{+} = 0 
\,,
\\
\left(\lambda_{00}(\omega)\,\omega^2 - 4\Delta^2\,[\lambda_{10}(\omega)-\lambda_{11}(\omega)]\right) 
E^{-} = 0 
\,,
\label{eq-Eminus}
\eer
where the functions $\lambda_{nm}(\omega)$ are moments of the $\vq=0$ Tsuneto function. 
For the anisotropic $\point{E}{1u}$ model with basis functions given in Eq. \ref{eq-E1u_basis} with
$\hat\vp_x=\cos\phi$ and $\hat\vp_y=\sin\phi$, we have for the anisotropic chiral ground state,
$\Delta(\vp)=\Delta\,e^{i\phi}\,I(\phi)$. The corresponding moments $\lambda_{nm}$ are then given
by
\be
\lambda_{nm}(\omega) 
= \Delta^2
\oint\frac{d\phi}{2\pi}\,\bar{\lambda}(\phi;\omega,\vq=0)\,\,[I(\phi)]^{2n}\,[\cos(2\phi)]^{2m}
\,.
\label{eq-Tsuneto_moments}
\ee
For the 2D chiral p-wave ground state the gap is isotropic on the Fermi circle, in which case 
$\lambda_{10}=\lambda_{00}=2\lambda_{11}=\lambda(\omega)$, 
with 
\be
\lambda(\omega) = \frac{1}{2}\abs{\Delta}^2
\int_{-\infty}^{+\infty}\frac{d\eps}{\sqrt{\eps^2 - \abs{\Delta}^2}}  
	\frac{\tanh\left(\frac{\beta\abs{\eps}}{2}\right)}{\eps^2 - (\omega/2)^2}
	\Theta\left(\varepsilon^2-\abs{\Delta}^2\right)
\,,
\label{eq-Tsuneto0}
\ee
leading to the degenerate $E^{\pm}$
modes with $M_{-,+}=M_{-,-}=\sqrt{2}\Delta(T)$, in agreement with the weak-coupling
limit of the TDGL theory; however now valid at any temperature.

The effects of anisotropy of the pairing state on the $E^{\pm}$ modes are computed by solving the
eigenvalue equations, Eqs. \ref{eq-Eplus},\ref{eq-Eminus} and \ref{eq-Tsuneto_moments} numerically.
The minimum and maximum of the anisotropic gap function are shown as a function of the anisotropy
parameter $\varepsilon$ in Fig. \ref{fig-Gap-EMode_Anisotropy}; $2\Delta_{\text{min}}$ marks
the minimum in the continuum edge of un-bound Fermion pairs.
The degeneracy of the $E^{\pm}$ modes is resolved by the anisotropy, however the splitting of the
modes for $T\rightarrow 0$ and relatively weak anisotropy, $\varepsilon \lesssim 0.4$, is much
smaller than the prediction based on the GL $\beta$ parameters.
The splitting of the two modes is generally smaller at lower temperatures, and becomes strongly
suppressed by the asymmetry in level repulsion between the $E^{\pm}$ modes and the continuum edge at
$2\Delta_{\text{min}}$ as the higher energy mode approaches the continuum edge.

At sufficiently large anisotropy the continuum edge of un-broken Fermion pairs at
$2\Delta_{\text{min}}$ intercepts the excitation energy of $E^{\pm}$ modes. This opens a channel for
the $E^{\pm}$ mode to dissociate into un-bound Fermion pairs, and thus leads to an intrinsic
lifetime for the $E^{\pm}$ mode(s), $\tau^{\pm}=\hbar/\Gamma^{\pm}$, where $\Gamma^{\pm}$ is the
width of the $E^{\pm}$ resonance.
The latter are calculated perturbatively from Eqs. \ref{eq-Eplus},\ref{eq-Eminus} and
\ref{eq-Tsuneto_moments} and are shown in the right panel of 
Fig. \ref{fig-Gap-EMode_Anisotropy}, onsetting precisely at an
anisotropy such that $M_{-,\pm}=2\Delta_{\text{min}}$. Note also that close to
$2\Delta_{\text{min}}$ the asymmetry in the level repulsion drives
$M_{-,-}\rightarrow M_{-,+}$. The large asymmetry in $\Gamma^{\pm}$ reflects the different phase space for
pair dissociation of the $E^{\pm}$ modes governed by
$\Im\lambda_{10}(\omega)\gg\Im\lambda_{11}(\omega)$, i.e the former is an isotropic average over the
spectrum of un-bound Fermion pairs, whereas the latter preferentially weights regions of the Fermi
surface near the gap maximum.
Thus, two key results of a self-consistent theory of coupled Boson-Fermion excitations are: (i) the
mass spliting of the $E^{\pm}$ mode spectrum is strongly suppressed by the asymmetry in the level
repulsion from the un-bound Fermion pairs, and (ii) there is a large asymmetry in the lifetimes of
the $E^{\pm}$ modes that results from the different phase space available for dissociation into
un-bound Fermion pairs by $E^{\pm}$ Bosons. Neither of these effects could
be anticipated a priori from the TDGL theory for the Bosonic excitations.

\vspace*{-3mm}
\subsection*{Microwave Excitation of the $E^{\pm}$ modes}
\vspace*{-3mm}

\begin{figure}[t]
\hspace*{-8.5cm}\begin{overpic}[scale=0.475]{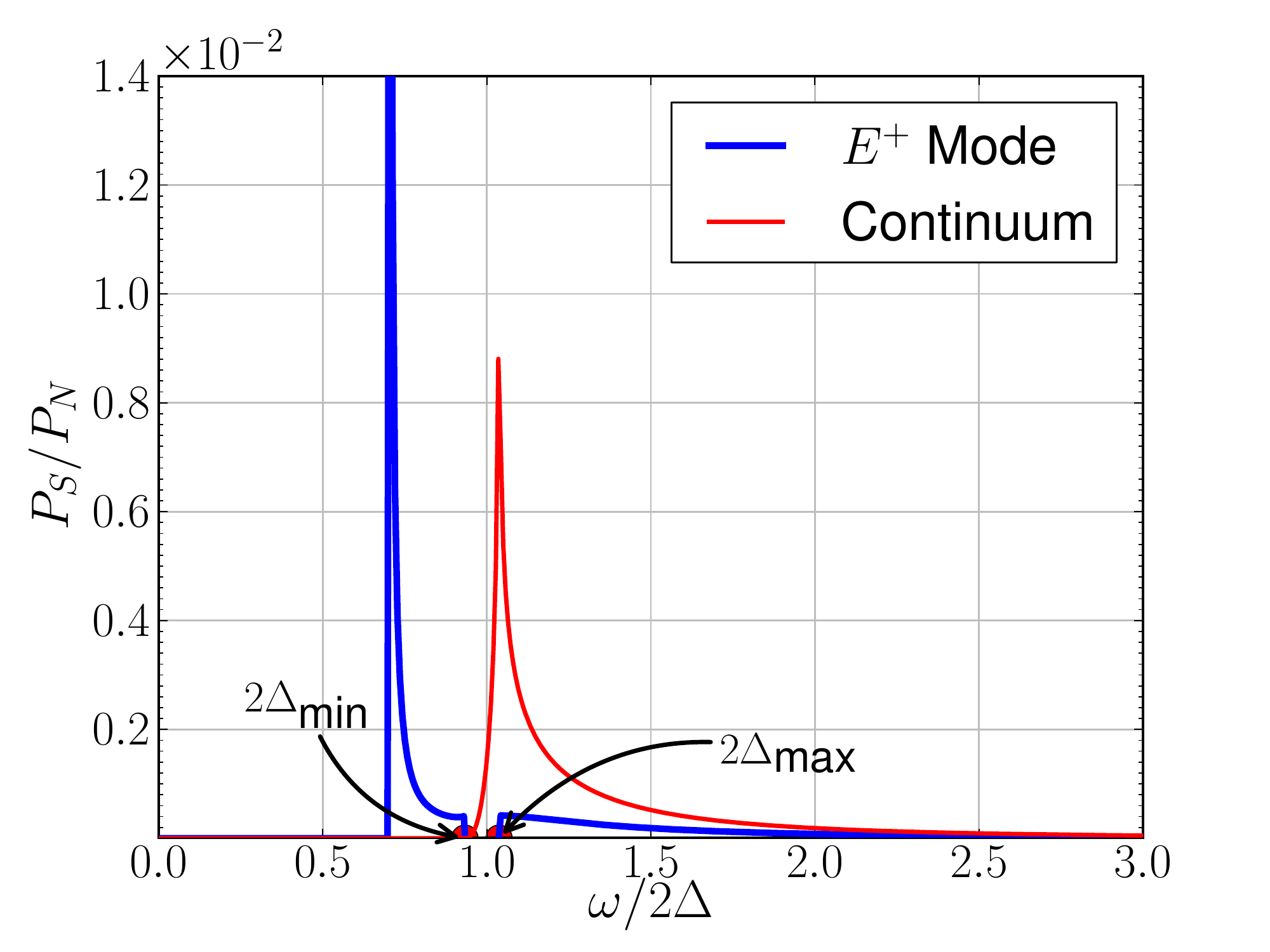}
\put(55,18){\includegraphics[scale=0.325]{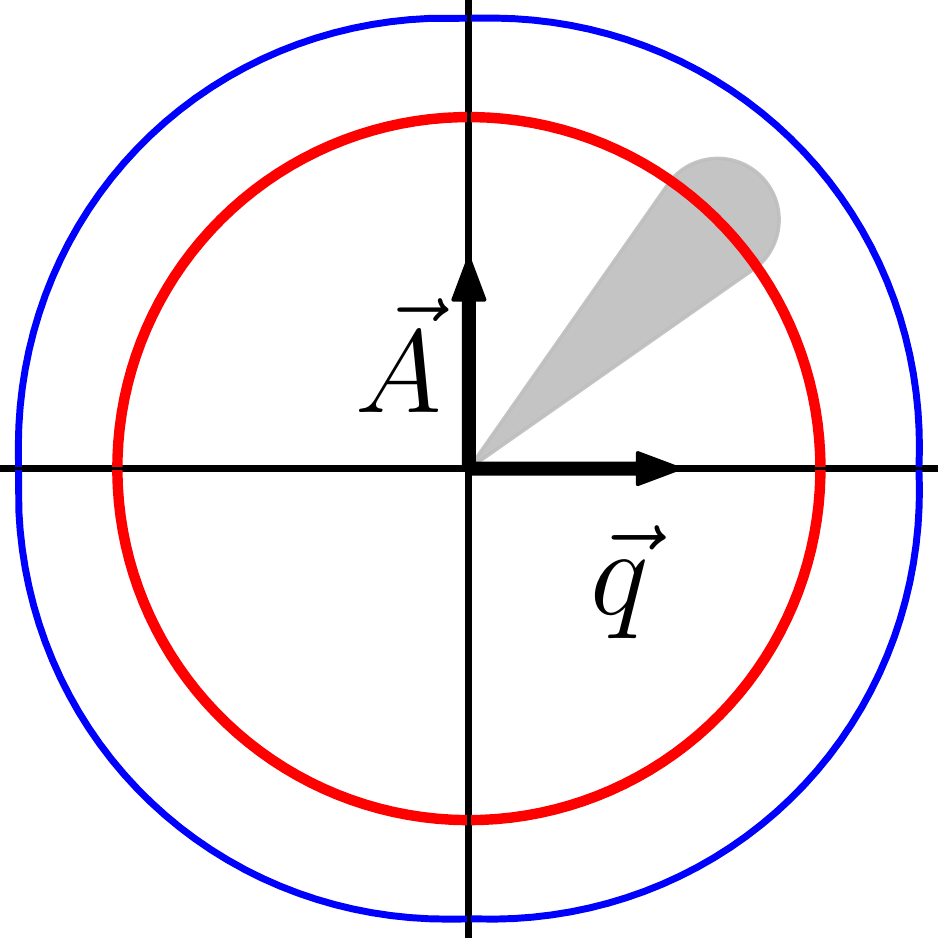}}
\put(95,0){\includegraphics[scale=0.475]{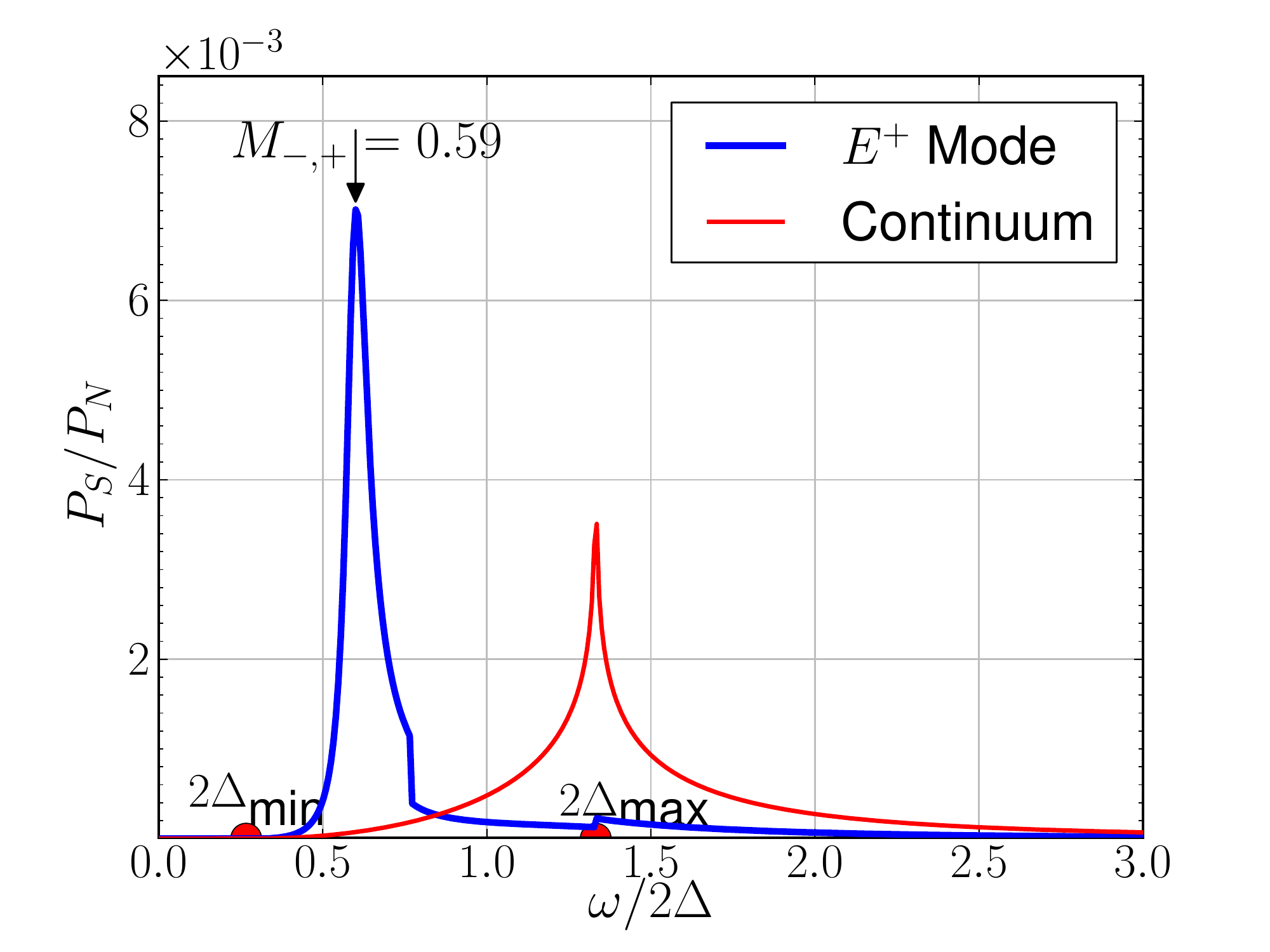}}
\put(145,18){\includegraphics[scale=0.35]{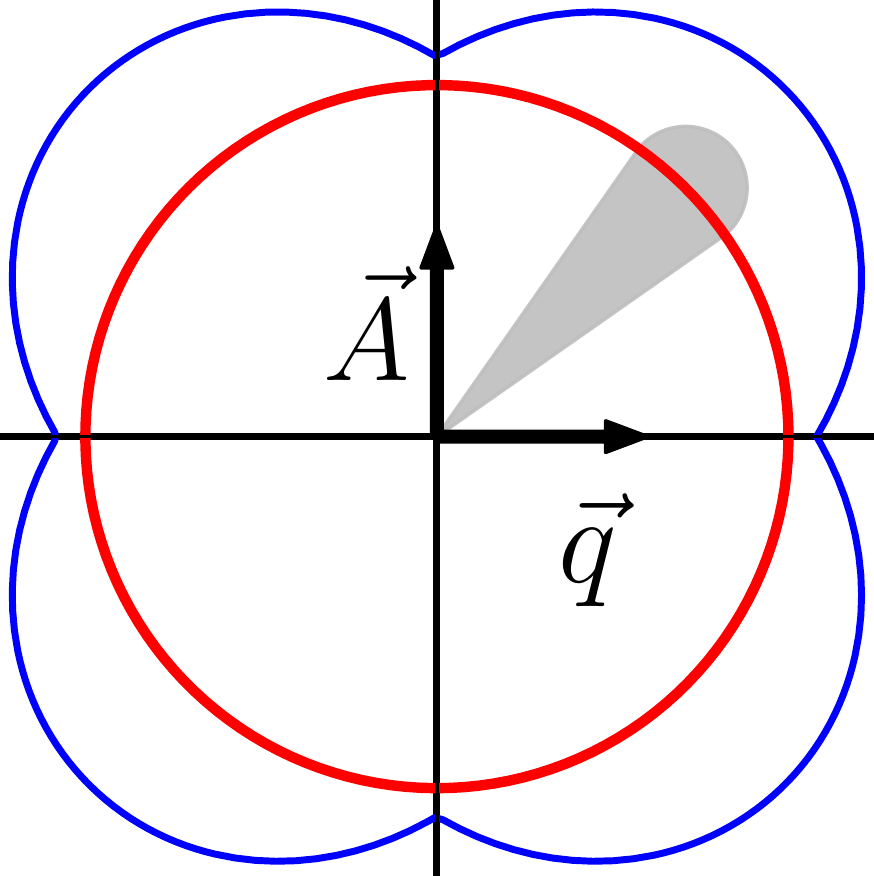}}
\end{overpic}
\caption{
Power absorption spectra normalized to the high-frequency limit of the normal-state,
$P_{\text{N}}(\omega)$, for $T=0$, penetration depth, $\Lambda/\xi=10$, and polarization
$\vq\perp\Aem$ along $[100]$ directions.
For this polarization direction only the $E^{+}$ mode is excited.
\underline{Left}: weak anisotropy with $\varepsilon=0.1$. A sharp $E^{+}$ absorption band (blue) and
a broad band of dissipation from pair dissociation that is sharply peaked at $2\Delta_{\text{max}}$
(red) are shown.
\underline{Right}: strong anisotropy with $\varepsilon=0.8$. A sharp $E^{+}$ absorption band (blue)
survives weak hybridization with the un-bound continuum. Dissipation from pair dissociation remains
sharply peaked at $2\Delta_{\text{max}}$.
} 
\label{fig-Power_Absorption_Eplus-110}
\end{figure}

Indeed key signatures of an anisotropic chiral ground state in \sro\, are the
excitation and decay channels for the $E^{\pm}$ modes. Both depend on the charge 
conjugation parity of the modes. 
Consider an EM field incident normal on a surface of \sro\ defined by the four-fold axis
of symmetry, $\hat\vz$, and an axis lying in the $x-y$ plane, i.e. $\vq\perp\hat\vz$, and with
linear polarization also in the $x-y$ plane, i.e. $\Aem\perp\hat\vz$ and $\Aem\perp\vq$. 
The EM field couples directly to the Fermionic degrees of freedom 
(particles and holes), generating a current \cite{hir89,wu15},
\be
\J_{\text{F}}(\vq,\omega) =  N_f\int\,d^2\vp\,(e\vv_{\vp})\,
	\left[1 + \frac{\eta^2}{\omega^2 - \eta^2}(1 - \lambda(\vp;\omega,\vq)) \right] 
	\left(\frac{e}{c}\vv_{\vp}\cdot\Aem\right)
\,.
\ee
Note that the effects of the pairing correlations 
on the Fermionic contribution to the charge current -
the opening of a gap in the Fermionic spectrum and the a.c. response of the negative energy
continuum (condensate) - are encoded in the Tsuneto function, $\lambda(\vp;\omega,\vq)$.
For $T\rightarrow 0$ and low frequencies, $\omega < 2\Delta_{\text{min}}$, only the negative energy
continuum (condensate) responds as an a.c. \emph{supercurrent} - $\pi/2$ out of phase with the
electric field - with zero dissipation.
The supercurrent, and thus the self-consistently determined EM field, are screened by the Meissner
effect and penetrate a distance of order the London penetration depth, $\Lambda$. This length scale
is typically large compared to the coherence length of the superconductor,
$\Lambda\gg\xi\gg\hbar/p_f$. In this limit the EM response is dominated by the bulk excitation
spectrum.
For a chiral ground state the EM field also couples directly to the $E^{\pm}$ Bosonic modes as shown
in Eqs. \ref{eq-Eplus}-\ref{eq-Eminus}. The Bosonic modes also generate a charge current,
\be
\J_{\text{B}}(\vq,\omega) =  \frac{1}{4}N_f\int\,d^2\vp\,(2e\vv_{\vp})\,
	(\vv_{\vp}\cdot\vq)\,\lambda(\vp;\omega,\vq))
	\left(\Delta^*(\vp)\,d(\vp;\vq,\omega) + \Delta(\vp)\,d^{'}(\vp;\vq,\omega)\right)
\,.
\ee
Thus, the total current, $\J=\J_{\text{F}}+\J_{\text{B}}$, can be expressed in terms of a response
function, $\J_i = \K_{ij}(\omega,\vq)\,\Aem_j(\vq,\omega)$, that encodes both Fermionic and Bosonic
contributions to the a.c. surface impedance. The response function $\K_{ij}(\vq,\omega)$ determines
both dissipative and non-dissipative contributions to the current. In particular, the microwave
power absorption spectrum can be expressed as a sum over the modes contributing to the Joule losses
of the electric field and current that penetrate into the superconductor, $P_{\text{S}}(\omega) =
-\frac{1}{2\pi}\int_{-\infty}^{+\infty}\,dq\, \Re[\J(q,\omega)\cdot\E^*(q,\omega)]$.
To calculate the power spectrum requires a solution of the boundary value problem for
the incident, reflected and transmitted EM fields. This boundary value problem determines
$\Aem(\vq,\omega)$ in terms of $\K_{ij}$ and the value of the EM field in vacuum at the surface,
$B_0$. For a detailed discussion of the boundary solution see Refs. \cite{yip92}.

For weak anisotropy, $\varepsilon=0.1$, the Bosonic modes have well defined
excitation energies, $\hbar\omega_{\pm}(\vq) = M_{-,\pm} + c_{\pm}^2|\vq|^2 / 2M_{-,\pm}$,
with $M_{-,\pm}\approx\sqrt{2}\Delta$ and $c_{\pm}\approx \frac{1}{2}v_f$, and carry current for
finite $\vq$ at frequencies below the un-bound Fermion pair continuum,
$\hbar\omega<2\Delta_{\text{min}}$, supported by the condensate, represented by
$\lambda(\vp;\vq,\omega)>0$.
Resonant excitation of the $E^{\pm}$ modes occurs over the frequency band,  
$M_{-,\pm}<\hbar\omega<2\Delta_{\text{min}}$, spanned by the dispersion of the $E^{\pm}$ modes.
Thus, excitation of the $E^{\pm}$ leads to an absorption band that is sharply peaked near threshold
as shown in the left panel of Fig. \ref{fig-Power_Absorption_Eplus-110} for $\varepsilon=0.1$ and
$E^{+}$. Note also, the broad band of dissipation from dissociation Cooper pairs into un-bound
Fermion pairs is sharply peaked at $2\Delta_{\text{max}}$.
For strong anisotropy, the $E^{\pm}$ modes broaden into resonances. However, the $E^{+}$ resonance
has a narrow linewidth due to limited phase space for decay into Fermion pairs. Furthermore, the
excitation of ground-state Cooper pairs into un-bound Fermion pairs is suppressed well below
$2\Delta_{\text{max}}$. Thus, even for strong gap suppression along $[100]$ directions there 
remains a strong absorption resonance from the weakly damped $E^{+}$ Bosonic mode.
Observation of the $E^{+}$ mode, and other signatures of the $E^{\pm}$ Bosonic spectrum, would provide direct evidence of an anisotropic chiral ground state in \sro.

\vspace*{-3mm}
\section*{Summary}
\vspace*{-3mm}

Chiral superconductors which break time-reversal symmetry necessarily belong to a higher dimensional
representation of the crystalline point group. In the cases of \sro, 2D \Hea, and \upt\ this is a
two-dimensional orbital representation.
An important consequence is that a chiral ground state supports Bosonic excitations of the
time-reversed Cooper pairs.
These excitations are degenerate for 2D \Hea\ with an excitation gap, $M=\sqrt{2}\Delta$, 
below the continuum edge of
un-bound Fermion pairs. 
Crystalline anisotropy lifts the degeneracy, and for strong anisotropy can generate a
low-lying Bosonic mode. Strong amplitude anisotropy also leads to low-lying Fermions, and thus a
channel for the Bosonic modes to decay in to un-bound Fermion pairs. 
Selection rules and phase space considerations are shown to generate to large asymmetries in the
lifetimes and hybridization of the Bosonic modes with the continuum of un-bound Fermion pairs.
The excitation of the Bosonic modes by microwave radiation could provide clear signatures of an 
anisotropic chiral ground state.
A detailed theory of microwave spectroscopy of anisotropic chiral superconductors which includes the
analysis of selection rules, the effects of band structure and Fermi surface anisotropy, spin-orbit
coupling and weak disorder will be forthcoming as a separate report \cite{wu15}.

\vspace*{-3mm}
\section*{Acknowledgements}
\vspace*{-3mm}
The research of HW and JAS was supported by the National Science Foundation (Grant DMR-1106315), 
while that of SBC is supported by the Institute for Basic Science of Korea (Grant IBS-R009-Y1).
JAS and SBC acknowledge the hospitality of the Aspen Center for Physics, and its support through
National Science Foundation Grant No. PHYS-1066293, where part of this work was carried out.
JAS also acknowledges the hospitality of the KITP and its support through NSF Grant No. PHY11-25915.
We acknowledge discussions with Sri Raghu, Catherine Kallin, Steve Simon and Thomas Scaffidi that
were important in motivating this work, and Andrea Damascelli for his comments on the role of 
spin-orbit coupling in \sro.
%

\end{document}